О. В. ДУДНИК, О. В. ЯКОВЛЄВ

Радіоастрономічний інститут НАН України,
вул. Мистецтв, 4, м. Харків, 61002, Україна
E-mail: dudnik@rian.kharkov.ua

# ВИЯВЛЕННЯ ВНУТРІШНІХ РАДІАЦІЙНИХ ПОЯСІВ ЗЕМЛІ У ПЕРІОД НИЗЬКОЇ СОНЯЧНОЇ І ГЕОМАГНІТНОЇ АКТИВНОСТІ ЗА ДАНИМИ ПРИЛАДУ СТЕП-Ф

Предмет і мета роботи: *Предметом досліджень є просторово-часові розподіли високоенергійних заряджених частинок усередині магнітосфери Землі поза зоною Південно-Атлантичної магнітної аномалії у період мінімуму 11-річного циклу сонячної активності. Метою роботи є пошуки і визначення сталих та нестійких додаткових просторових зон підвищених потоків електронів субрелятивістських енергій на висотах польотів низькоорбітальних штучних супутників Землі.*

Методи і методологія: *Пошук і встановлення додаткових радіаційних поясів Землі здійснено за аналізом даних каналу D1e реєстрації електронів з енергіями* $\Delta E_e = 180 \div 510$ *кеВ і протонів з енергіями* $\Delta E_p = 3.5 \div 3.7$ *МеВ супутникового телескопу електронів і протонів СТЕП-Ф на борту низькоорбітального наукового супутника Землі "КОРОНАС-Фотон". Для аналізу використовувались інформаційні масиви з часовою роздільною здатністю 2 с, нормовані на активну площу позиційно-чутливого кремнієвого матричного детектора та тілесний кут зору детекторної голівки приладу.*

Результати: *Виявлено сталу структуру з трьох електронних радіаційних поясів у земній магнітосфері в період низької сонячної і геомагнітної активності у травні 2009 р. Два пояси є відомими з початку космічної ери радіаційними поясами Ван Алена, ще один додатковий сталий шар формується навколо дрейфової оболонки з параметром Мак-Ілвайна* $L = 1.65 \pm 0.05$. *В окремі дні травня 2009 р., крім цього додаткового шару, одночасно з ним спостерігався ще один нестійкий внутрішній радіаційний пояс, що формувався час від часу між досліджуваним сталим поясом на* $L \approx 1.65$ *і внутрішнім поясом Ван Алена на* $L \approx 2.52$. *Підвищені потоки частинок у цьому нестійкому поясі формувалися навколо дрейфової оболонки* $L = 2.06 \pm 0.14$.

Висновки: *Додаткові внутрішні радіаційні пояси реєструються в широкій смузі географічних довгот* $\lambda$ *як на висхідних, так і на низхідних ділянках орбіти супутника, від* $\lambda_1 \approx 150^\circ$ *до* $\lambda_2 \approx 290^\circ$. *Окремо у Північній або Південній півкулі поза межами зовнішнього краю зовнішнього радіаційного поясу на* $L \geq 7 \div 8$ *спостерігаються випадки підвищення щільності потоку частинок в широкій смузі L-оболонок, які відповідають високоширотній області квазізахвату високоенергійних заряджених частинок. Підвищені потоки спостерігаються аж до меж головної ударної хвилі земної магнітосфери (* $L \approx 10 \div 12$*).*

## 1. Вступ

Просторові зони внутрішнього і зовнішнього електронних радіаційних поясів Землі є вельми дослідженими впродовж останніх десятиріч завдяки обробці та аналізу численних експериментальних даних, отриманих штучними супутниками Землі. Пітч-кутова і радіальна дифузії електронів, розщеплення дрейфових оболонок геомагнітного поля та потоків високоенергійних іонів і електронів, дрейфове відлуння та інші ефекти, пов'язані з впливом геомагнітних бур, корональних викидів маси, високошвидкісних потоків сонячного вітру та міжпланетних ударних хвиль на магнітосферу Землі, широко представлені в літературі [1–5].

Ще на початку активних досліджень було показано, що найбільш вагомими типами висипань високоенергійних частинок є висипання електронів з внутрішнього поясу, а також протонів з кільцевого струму [1]. Електронні висипання виникають через резонансну взаємодію типу "хвиля–частинка", тоді як протони висипаються до низьких висот з кільцевого струму через обмін зарядами з воднем екзосфери, перетворюючись на нейтрали. Геомагнітні бурі можуть збільшувати або зменшувати потоки релятивістських електронів в радіаційних поясах. Через статистичний аналіз впливу геомагнітних бур у роботі [6] доведено, що приблизно половина з 276 досліджуваних бур збільшувала потоки, кожна п'ята буря зменшувала потоки, а решта бур (28 %) викликала зміни, які збільшували або зменшували попереднє значення потоків не більш, ніж у два рази.

Менш дослідженими є явища спорадичного виникнення додаткових зон підвищеної радіації у проміжку між зовнішнім і внутрішнім поясами Ван Алена на низьких широтах. Моноенергетичні сплески електронів спостерігались у конусі втрат на дрейфових оболонках, що характеризуються значеннями параметра Мак-Ілвайна $L = 1.6 \div 1.85,$ в діапазоні східних довгот від $\lambda = 55°$ до $\lambda = 62°$ [7]. Позиціювання $L$-оболонки електронних висипань співпадало з місцем розташування наземної потужної радіопередавальної низькочастотної станції, що вказувало на вплив радіовипромінювання на захоплені в радіаційних поясах електрони.

У роботах [8, 9] показано, що поза межами Південно-Атлантичної аномалії (ПАА) і авроральної зони чітко відслідковується підвищення потоків електронів з енергіями $E_e = 0.19 \div 3.2$ МеВ у вузькій смузі широт (близько 3°) на середніх широтах над Австралією і Мексикою. Довжина цих середньоширотних зон підвищених потоків на $L = 1.6$ складає кілька десятків градусів за довготами в обох півкулях. Усереднений пік локалізації електронів і протонів з енергіями $E_p = 0.58 \div 35$ МеВ знаходиться на $L = 1.71,$ $\Delta L = \pm 0.16$ [9]. Вимірювальна апаратура на супутникові CRRES зареєструвала 24 березня 1991 р. майже миттєву інжекцію електронів і протонів з енергіями вище 15 МеВ у прошарок $L$-оболонок $2 \le L \le 3$ [3, 4]. У роботі [10] наголошено, що додаткові радіаційні пояси як протонів, так і електронів можуть існувати протягом тривалих періодів часу, від місяців до років, після суттєвих інжекцій високоенергійних частинок вглиб магнітосфери.

Метою роботи є пошук і встановлення сталих і нестійких додаткових просторових зон підвищених потоків електронів проміжних енергій за даними низькоенергетичного каналу реєстрації частинок супутниковим телескопом електронів і протонів СТЕП-Ф на борту низькоорбітального наукового штучного супутника Землі “КОРОНАС-Фотон”.

## 2. Принципи відбору даних для аналізу

Попередній аналіз отриманої інформації показав, що у низькоенергетичному змішаному каналі реєстрації електронів і протонів D1e приладу СТЕП-Ф [11–13] регулярно спостерігалась структура з трьох радіаційних поясів у період низької геомагнітної і сонячної активності в травні 2009 р. [14, 15]. У зв’язку з цим до бази даних, створеної з метою поглибленого аналізу наукової інформації, не брались до уваги дані, накопичені під час прольотів космічного апарата над зоною ПАА. Масиви наукових даних, які містили в собі періоди детектування відомих електронних радіаційних поясів Ван Алена, розглядались як підставові для порівняльного аналізу. Відповідно, розглядалися просторові зони і періоди реєстрації двох відомих і додаткових, досі ще не вивчених, електронних радіаційних поясів або ж нестабільних, виникаючих на певний проміжок часу, підвищених потоків частинок. В процесі попереднього огляду масивів даних було з’ясовано, що додатковий радіаційний пояс, умовно позначений як $S_2/N_2$ ($S_2$ – у Південній півкулі, $N_2$ – у Північній), може спостерігатись навіть за відсутності потоків у внутрішньому радіаційному поясі Ван Алена. Тому такі випадки також додавались до бази даних для загального аналізу.

## 3. Виявлення тришарової структури електронних радіаційних поясів у період низької сонячної і геомагнітної активності

Огляд інформаційних масивів зі змішаного каналу D1e реєстрації електронів з енергіями $\Delta E_e = 180 \div 510$ кеВ і протонів з енергіями $\Delta E_p = 3.5 \div 3.7$ МеВ з часовою роздільною здатністю даних $\tau = 2$ с показав, що окрім підвищених потоків у радіаційних поясах Ван Алена і в зоні ПАА, існує щонайменше ще один шар з додатковим внутрішнім електронним радіаційним поясом. Ці потоки формуються навколо оболонки $L \approx 1.65.$ Цей шар спостерігався під час лише декількох з 15 добових орбітальних обертів супутника. Кількість таких обертів складала від 1 до 6, причому додатковий сталий пояс спостерігався як на висхідних, так і на низхідних ділянках орбіти космічного апарата.

Зокрема, у травні 2009 р. чітка структура з трьох електронних радіаційних поясів спостерігалась щоденно, переважно у нічні передранкові часи за місцевим часом. Для прикладу в табл. 1 наведено результати спостережень додаткового

*Таблиця 1.* **Результати спостережень додаткового сталого внутрішнього електронного радіаційного поясу Землі $S_2/N_2$ у період з 1 по 31 травня 2009 р.**

| Дата | Номер оберту супутника | Час реєстрації поясу $S_2$ у Південній півкулі, год:хв:с | | *L*-оболонка | D1e, частинка/(см²·с·ср) | Довгота, ° (0°÷360°) | Широта, ° (–90°÷90°) | Час реєстрації поясу $N_2$ у Північній півкулі, год:хв:с | | *L*-оболонка | D1e, частинка/(см²·с·ср) | Довгота, ° (0°÷360°) | Широта, ° (–90°÷90°) |
|---|---|---|---|---|---|---|---|---|---|---|---|---|---|
| | | Місцевий (LT) | Світовий (UT) | | | | | Місцевий (LT) | Світовий (UT) | | | | |
| 01 | 1 | – | – | – | – | – | – | 03:41:16 | 11:55:16 | 1.6 | 35 | 237 | 30 |
| | 2 | 03:01:21 | 13:13:16 | 1.7 | 64 | 207 | –37 | 03:45:42 | 13:32:46 | 1.7 | 207 | 213 | 36 |
| | 3 | 03:03:50 | 14:50:02 | 1.7 | 13 | 184 | –33 | 03:49:24 | 15:09:48 | 1.7 | 163 | 190 | 41 |
| | 4 | – | – | – | – | – | – | 03:52:34 | 16:46:32 | 1.7 | 29 | 167 | 45 |
| | 5 | 15:54:12 | 22:22:24 | 1.7 | 45 | 263 | –48 | – | – | – | – | – | – |
| 02 | 6 | 02:52:53 | 11:33:02 | 1.5 | 55 | 230 | –38 | 03:33:16 | 11:51:02 | 1.6 | 116 | 236 | 29 |
| | 7 | 02:53:12 | 13:09:02 | 1.7 | 17 | 206 | –37 | 03:37:42 | 13:28:34 | 1.7 | 97 | 212 | 36 |
| 03 | 8 | 02:43:11 | 11:28:18 | 1.6 | 41 | 229 | –40 | 03:26:17 | 11:47:14 | 1.6 | 50 | 235 | 31 |
| | 9 | 02:46:04 | 13:05:06 | 1.7 | 30 | 205 | –36 | 03:30:13 | 13:24:32 | 1.7 | 70 | 211 | 37 |
| | 10 | – | – | – | – | – | – | 03:34:30 | 15:01:44 | 1.7 | 40 | 188 | 42 |
| | 11 | – | – | – | – | – | – | 03:36:56 | 16:38:16 | 1.7 | 27 | 165 | 45 |
| | 12 | 15:39:32 | 22:14:20 | 1.8 | 69 | 261 | –49 | – | – | – | – | – | – |
| 04 | 13 | 02:35:49 | 11:24:20 | 1.6 | 149 | 228 | –40 | 03:18:32 | 11:43:10 | 1.6 | 157 | 234 | 31 |
| | 14 | 02:38:33 | 13:01:06 | 1.6 | 39 | 204 | –36 | 03:21:59 | 13:20:18 | 1.6 | 166 | 210 | 36 |
| | 15 | – | – | – | – | – | – | 03:25:51 | 14:57:24 | 1.7 | 34 | 187 | 41 |
| | 16 | 15:28:51 | 22:09:34 | 1.7 | 42 | 260 | –46 | – | – | – | – | – | – |
| | 17 | 15:22:57 | 23:43:52 | 1.6 | 28 | 235 | –41 | 14:37:14 | 23:24:04 | 1.7 | 91 | 228 | 33 |
| 05 | 18 | 02:28:57 | 11:20:32 | 1.6 | 245 | 227 | –39 | 03:10:38 | 11:39:02 | 1.6 | 210 | 233 | 31 |
| | 19 | 02:30:48 | 12:57:02 | 1.6 | 68 | 203 | –36 | 03:14:21 | 13:16:16 | 1.6 | 204 | 210 | 36 |
| | 20 | – | – | – | – | – | – | 03:18:22 | 14:53:24 | 1.7 | 38 | 186 | 42 |
| | 21 | 15:21:59 | 22:05:42 | 1.7 | 43 | 259 | –47 | – | – | – | – | – | – |
| | 22 | – | – | – | – | – | – | 14:28:11 | 23:19:32 | 1.7 | 53 | 227 | 35 |
| 06 | 23 | 02:21:08 | 11:16:20 | 1.6 | 123 | 226 | –39 | 03:02:48 | 11:34:50 | 1.6 | 241 | 241 | 31 |
| | 24 | 02:23:36 | 12:53:02 | 1.6 | 92 | 203 | –35 | 03:06:53 | 13:12:12 | 1.7 | 283 | 209 | 37 |
| | 25 | 02:25:50 | 14:29:46 | 1.6 | 15 | 179 | –31 | 03:11:12 | 14:49:24 | 1.7 | 44 | 186 | 42 |
| | 26 | – | – | – | – | – | – | 03:13:12 | 16:25:48 | 1.7 | 40 | 162 | 45 |
| 07 | 27 | – | – | – | – | – | – | 02:55:23 | 11:30:54 | 1.6 | 68 | 231 | 31 |
| | 28 | 02:15:15 | 12:48:46 | 1.7 | 217 | 202 | –36 | 02:59:11 | 13:08:08 | 1.7 | 803 | 208 | 37 |
| | 29 | 02:16:45 | 14:25:12 | 1.7 | 16 | 178 | –33 | 03:03:46 | 14:45:24 | 1.7 | 74 | 185 | 43 |
| 08 | 30 | 02:07:27 | 11:08:46 | 1.5 | 42 | 225 | –36 | 02:46:12 | 11:26:16 | 1.5 | 57 | 230 | 29 |
| | 31 | 02:07:40 | 12:44:44 | 1.7 | 111 | 201 | –35 | 02:51:40 | 13:04:08 | 1.7 | 273 | 207 | 37 |
| | 32 | 02:09:45 | 14:21:24 | 1.7 | 18 | 177 | –32 | 02:55:00 | 14:41:02 | 1.7 | 137 | 184 | 42 |

*Таблиця 1*. Продовження

| Дата | Номер оберту супутника | Час реєстрації поясу $S_2$ у Південній півкулі, год:хв:с | | *L*-оболонка | D1e, частинка/ (см$^2$·с·ср) | Довгота, ° (0°÷360°) | Широта, ° (–90°÷90°) | Час реєстрації поясу $N_2$ у Північній півкулі, год:хв:с | | *L*-оболонка | D1e, частинка/ (см$^2$·с·ср) | Довгота, ° (0°÷360°) | Широта, ° (–90°÷90°) |
|---|---|---|---|---|---|---|---|---|---|---|---|---|---|
| | | Місцевий (LT) | Світовий (UT) | | | | | Місцевий (LT) | Світовий (UT) | | | | |
| 09 | 33 | – | – | – | – | – | – | 02:48:42 | 14:37:22 | 1.7 | 96 | 183 | 43 |
| | 34 | 14:51:52 | 21:49:28 | 1.8 | 165 | 256 | –48 | 13:59:04 | 21:27:54 | 1.8 | 56 | 248 | 33 |
| | 35 | 14:45:32 | 23:23:44 | 1.7 | 36 | 231 | –43 | 13:55:54 | 23:02:42 | 1.8 | 285 | 223 | 36 |
| 10 | 36 | 01:52:12 | 12:36:36 | 1.7 | 51 | 199 | –35 | 02:35:30 | 12:55:46 | 1.6 | 119 | 205 | 37 |
| | 37 | – | – | – | – | – | – | 02:39:33 | 14:32:54 | 1.7 | 67 | 182 | 42 |
| | 38 | 14:35:23 | 23:18:58 | 1.6 | 37 | 229 | –40 | 13:49:10 | 22:58:58 | 1.7 | 82 | 223 | 35 |
| 11 | 39 | 01:42:02 | 10:55:50 | 1.6 | 76 | 222 | –39 | 02:24:13 | 11:14:32 | 1.6 | 210 | 227 | 31 |
| | 40 | 01:44:05 | 12:32:24 | 1.7 | 93 | 198 | –35 | 02:28:12 | 12:51:50 | 1.6 | 319 | 204 | 37 |
| | 41 | 01:45:33 | 14:08:50 | 1.8 | 19 | 174 | –33 | 02:32:06 | 14:28:54 | 1.7 | 75 | 181 | 42 |
| | 42 | 14:37:19 | 20:05:40 | 1.7 | 670 | 278 | –9 | – | – | – | – | – | – |
| | 43 | 14:33:40 | 21:40:40 | 1.7 | 75 | 253 | –46 | 13:43:59 | 21:19:54 | 1.8 | 49 | 246 | 32 |
| | 44 | 14:28:07 | 23:15:02 | 1.7 | 232 | 228 | –41 | 13:40:20 | 22:54:32 | 1.7 | 593 | 222 | 36 |
| 12 | 45 | 01:34:12 | 10:51:38 | 1.6 | 48 | 221 | –39 | 02:16:32 | 11:10:24 | 1.6 | 134 | 227 | 32 |
| | 46 | 01:36:33 | 12:28:18 | 1.7 | 184 | 197 | –35 | 02:20:33 | 12:47:42 | 1.6 | 521 | 203 | 37 |
| | 47 | 02:24:36 | – | – | – | – | – | 02:24:36 | 14:24:48 | 1.7 | 40 | 180 | 43 |
| 13 | 48 | 01:28:07 | 10:48:02 | 1.6 | 68 | 220 | –37 | 02:08:07 | 11:06:02 | 1.6 | 129 | 226 | 31 |
| | 49 | 01:29:22 | 12:24:22 | 1.6 | 89 | 196 | –34 | 02:12:42 | 12:43:32 | 1.6 | 286 | 202 | 37 |
| | 50 | – | – | – | – | – | – | 02:16:36 | 14:20:36 | 1.7 | 51 | 179 | 42 |
| | 51 | 14:18:30 | 21:32:22 | 1.7 | 39 | 252 | –46 | – | – | – | – | – | – |
| | 52 | 14:12:11 | 23:06:32 | 1.7 | 71 | 226 | –41 | 13:25:09 | 22:46:16 | 1.7 | 193 | 220 | 35 |
| 14 | 53 | 01:21:32 | 12:20:12 | 1.6 | 40 | 195 | –34 | 02:04:56 | 12:39:24 | 1.6 | 121 | 201 | 37 |
| | 54 | 01:23:23 | 13:56:48 | 1.7 | 17 | 172 | –31 | 02:09:05 | 14:16:32 | 1.7 | 131 | 178 | 43 |
| | 55 | 14:13:13 | 19:52:58 | 1.7 | 357 | 275 | –48 | – | – | – | – | – | – |
| | 56 | 14:03:12 | 23:02:02 | 1.6 | 57 | 225 | –9 | 13:18:10 | 22:42:24 | 1.6 | 93 | 219 | 34 |
| 15 | 57 | – | – | – | – | – | – | 13:14:19 | 00:17:02 | 1.6 | 194 | 194 | 38 |
| | 58 | 01:12:59 | 10:39:54 | 1.5 | 62 | 218 | –36 | 01:53:24 | 10:58:04 | 1.6 | 112 | 224 | 32 |
| | 59 | 01:14:19 | 12:16:16 | 1.6 | 123 | 195 | –33 | 01:57:16 | 12:35:18 | 1.6 | 274 | 201 | 38 |
| | 60 | – | – | – | – | – | – | 02:00:58 | 14:12:18 | 1.7 | 84 | 177 | 42 |
| | 61 | 13:56:41 | 22:58:16 | 1.7 | 225 | 225 | –41 | 13:09:00 | 22:37:48 | 1.7 | 234 | 218 | 36 |
| 16 | 62 | 01:03:36 | 10:35:14 | 1.6 | 127 | 217 | –38 | 01:45:55 | 10:54:02 | 1.6 | 145 | 223 | 32 |
| | 63 | 01:06:33 | 12:12:08 | 1.6 | 215 | 194 | –33 | 01:49:57 | 12:31:18 | 1.6 | 1050 | 200 | 38 |
| | 64 | 01:08:02 | 13:48:36 | 1.7 | 20 | 170 | –31 | 01:53:51 | 14:08:20 | 1.7 | 189 | 176 | 43 |
| | 65 | 13:49:37 | 22:54:20 | 1.7 | 114 | 224 | –42 | 13:00:22 | 22:33:22 | 1.7 | 309 | 217 | 37 |

*Таблиця 1.* Продовження

| Дата | Номер оберту супутника | Час реєстрації поясу $S_2$ у Південній півкулі, год:хв:с | | *L*-оболонка | D1e, частинка/ (см²·с·ср) | Довгота, ° (0°÷360°) | Широта, ° (–90°÷90°) | Час реєстрації поясу $N_2$ у Північній півкулі, год:хв:с | | *L*-оболонка | D1e, частинка/ (см²·с·ср) | Довгота, ° (0°÷360°) | Широта, ° (–90°÷90°) |
|---|---|---|---|---|---|---|---|---|---|---|---|---|---|
| | | Місцевий (LT) | Світовий (UT) | | | | | Місцевий (LT) | Світовий (UT) | | | | |
| 17 | 66 | 00:55:50 | 10:31:06 | 1.6 | 102 | 216 | –38 | 01:38:54 | 10:50:10 | 1.6 | 195 | 222 | 34 |
| | 67 | 00:58:57 | 12:08:04 | 1.6 | 131 | 193 | –33 | 01:42:24 | 12:27:14 | 1.7 | 302 | 199 | 39 |
| | 68 | – | – | – | – | – | – | 01:46:20 | 14:04:16 | 1.7 | 42 | 176 | 43 |
| 18 | 69 | 00:49:54 | 10:27:34 | 1.5 | 68 | 216 | –35 | 01:29:56 | 10:45:36 | 1.6 | 115 | 221 | 32 |
| | 70 | 00:51:29 | 12:04:02 | 1.6 | 92 | 192 | –33 | 01:34:24 | 12:23:02 | 1.6 | 229 | 198 | 38 |
| | 71 | – | – | – | – | – | – | 01:38:50 | 14:00:12 | 1.7 | 69 | 175 | 44 |
| | 72 | 13:42:45 | 19:36:32 | 1.7 | 88 | 272 | –49 | – | – | – | – | – | – |
| | 73 | 13:32:32 | 22:45:36 | 1.7 | 230 | 222 | –0 | 12:45:44 | 22:25:24 | 1.7 | 422 | 215 | 36 |
| 19 | 74 | 00:41:04 | 10:23:16 | 1.6 | 51 | 215 | –37 | 01:21:56 | 10:41:34 | 1.6 | 107 | 220 | 32 |
| | 75 | 00:43:40 | 12:00:04 | 1.6 | 53 | 191 | –33 | 01:26:31 | 12:19:02 | 1.6 | 121 | 197 | 38 |
| | 76 | 13:24:41 | 22:41:36 | 1.7 | 49 | 221 | –40 | 12:36:26 | 22:20:54 | 1.7 | 125 | 214 | 38 |
| 20 | 77 | 00:34:47 | 11:55:32 | 1.7 | 50 | 190 | –34 | 01:19:30 | 12:15:08 | 1.7 | 94 | 196 | 39 |
| | 78 | 00:37:01 | 13:32:16 | 1.7 | 24 | 166 | –30 | 01:22:34 | 13:51:54 | 1.7 | 83 | 173 | 43 |
| | 79 | 13:27:12 | 19:28:24 | 1.7 | 346 | 270 | –49 | – | – | – | – | – | – |
| | 80 | 13:22:55 | 21:03:14 | 1.7 | 49 | 245 | –46 | – | – | – | – | – | – |
| | 81 | 13:16:26 | 22:37:18 | 1.7 | 65 | 220 | –40 | 12:29:36 | 22:17:04 | 1.7 | 99 | 213 | 36 |
| 21 | 82 | 00:24:54 | 10:14:46 | 1.6 | 47 | 213 | –37 | 01:07:55 | 10:33:50 | 1.6 | 118 | 219 | 34 |
| | 83 | 00:27:59 | 11:51:44 | 1.6 | 60 | 189 | –33 | 01:10:47 | 12:10:42 | 1.6 | 194 | 195 | 38 |
| | 84 | 00:29:26 | 13:28:12 | 1.7 | 38 | 165 | –30 | 01:14:37 | 13:47:42 | 1.7 | 267 | 172 | 43 |
| 22 | 85 | 00:18:01 | 10:10:56 | 1.6 | 48 | 212 | –36 | 00:59:58 | 10:29:38 | 1.6 | 192 | 218 | 34 |
| | 86 | 00:20:35 | 11:47:44 | 1.6 | 47 | 188 | –32 | 01:03:24 | 12:06:40 | 1.6 | 183 | 194 | 39 |
| | 87 | 00:21:51 | 13:24:08 | 1.7 | 23 | 164 | –30 | 01:07:20 | 13:43:42 | 1.7 | 80 | 171 | 44 |
| 23 | 88 | 00:12:48 | 11:43:36 | 1.6 | 64 | 187 | –32 | 00:55:58 | 12:02:38 | 1.6 | 472 | 193 | 39 |
| | 89 | 00:14:24 | 13:20:08 | 1.6 | 27 | 164 | –29 | 00:59:22 | 13:39:30 | 1.7 | 61 | 170 | 43 |
| 24 | 90 | – | – | – | – | – | – | 00:43:47 | 10:21:08 | 1.6 | 74 | 216 | 33 |
| | 91 | 00:05:51 | 11:39:46 | 1.6 | 73 | 186 | –31 | 00:47:22 | 11:58:14 | 1.6 | 345 | 192 | 38 |
| 25 | 92 | 23:57:52 | 11:35:32 | 1.6 | 28 | 186 | –31 | 00:41:15 | 11:54:36 | 1.7 | 72 | 192 | 40 |
| | 93 | 23:59:38 | 13:12:10 | 1.6 | 34 | 162 | –28 | 00:43:09 | 13:31:02 | 1.6 | 144 | 168 | 43 |
| | 94 | 12:47:51 | 19:07:34 | 1.7 | 130 | 265 | –49 | – | – | – | – | – | – |
| 26 | 95 | 23:49:34 | 11:31:24 | 1.6 | 28 | 185 | –32 | 00:32:14 | 11:50:16 | 1.6 | 111 | 191 | 39 |
| | 96 | 23:50:46 | 13:07:46 | 1.7 | 22 | 161 | –29 | 00:35:45 | 13:27:10 | 1.6 | 61 | 167 | 43 |
| 27 | 97 | 23:41:35 | 11:27:10 | 1.6 | 215 | 184 | –32 | 00:24:26 | 11:46:06 | 1.6 | 756 | 190 | 39 |
| | 98 | 23:42:55 | 13:03:36 | 1.7 | 21 | 160 | –29 | 00:28:04 | 13:23:02 | 1.7 | 61 | 166 | 43 |

*Таблиця 1.* Продовження

| Дата | Номер оберту супутника | Час реєстрації поясу $S_2$ у Південній півкулі, год:хв:с | | *L*-оболонка | D1e, частинка/(см²·с·ср) | Довгота, ° (0°÷360°) | Широта, ° (–90°÷90°) | Час реєстрації поясу $N_2$ у Північній півкулі, год:хв:с | | *L*-оболонка | D1e, частинка/(см²·с·ср) | Довгота, ° (0°÷360°) | Широта, ° (–90°÷90°) |
|---|---|---|---|---|---|---|---|---|---|---|---|---|---|
| | | Місцевий (LT) | Світовий (UT) | | | | | Місцевий (LT) | Світовий (UT) | | | | |
| 28 | 99 | – | – | – | – | – | – | 00:17:29 | 11:42:12 | 1.6 | 86 | 189 | 40 |
| | 100 | 12:12:43 | 22:03:50 | 1.7 | 114 | 212 | –38 | 11:25:47 | 21:43:32 | 1.7 | 164 | 206 | 38 |
| | 101 | 12:08:28 | 23:38:18 | 1.7 | 23 | 188 | –34 | 11:21:33 | 23:18:06 | 1.7 | 73 | 181 | 42 |
| 29 | 102 | 23:26:31 | 11:19:02 | 1.6 | 57 | 182 | –31 | 00:09:34 | 11:38:00 | 1.6 | 237 | 188 | 40 |
| | 103 | 23:28:05 | 12:55:34 | 1.6 | 70 | 158 | –28 | 00:12:27 | 13:14:42 | 1.6 | 125 | 164 | 43 |
| 30 | 104 | 23:18:25 | 11:14:46 | 1.6 | 40 | 181 | –32 | 00:02:27 | 11:34:02 | 1.6 | 165 | 187 | 40 |
| 31 | 105 | – | – | – | – | – | – | 23:50:24 | 09:52:38 | 1.6 | 49 | 209 | 35 |
| | 106 | 23:11:30 | 11:10:56 | 1.6 | 24 | 180 | –30 | 23:53:50 | 11:29:38 | 1.6 | 130 | 186 | 40 |
| | 107 | 23:12:52 | 12:47:24 | 1.6 | 30 | 156 | –28 | 23:56:34 | 13:06:18 | 1.6 | 98 | 163 | 43 |
| | 108 | – | – | – | – | – | – | 10:58:27 | 23:05:44 | 1.6 | 60 | 178 | 42 |
| | | Усереднене значення *L* 1.65±0.05 | | | | | | Усереднене значення *L* 1.65±0.05 | | | | | |

сталого внутрішнього радіаційного поясу Землі $S_2/N_2$ у період з 1 по 31 травня 2009 р. з зазначенням дати, наскрізного номера оберту супутника, місцевого і світового часів, *L*-оболонки, на якій реєструвалося максимальне значення щільності потоку електронів, географічних координат і власне максимального значенням цього потоку. Прочерками позначені випадки відсутності підвищених (порівняно з оточуючими фоновими) значень потоків у додатковому радіаційному поясі в одній з півкуль.

З табл. 1 видно, що додатковий шар підвищених потоків електронів з енергіями $E_e = 0.18 \div 0.51$ МеВ спостерігався щоденно, попри спокійний характер або незначні варіації геомагнітного поля Землі, що мали місце у цей період [14]. Кількість обертів, під час яких був стабільно помітним додатковий пояс, варіювала від 1 до 6.

Максимальні потоки електронів у зовнішньому радіаційному поясі Ван Алена значно перевищували найбільші потоки у внутрішніх поясах. З метою порівняльного аналізу розподілу щільності потоку частинок на всіх *L*-оболонках, починаючи від відкритих силових ліній геомагнітного поля до найменших значень *L*-параметра, до загальної картини було долучено часовий хід щільності потоку частинок у зовнішньому радіаційному поясі.

На рис. 1 наведено типовий приклад реєстрації трьох електронних радіаційних поясів в обох півкулях у травні 2009 р. вночі за місцевим часом на висхідній від Південної до Північної півкулі ділянці орбіти супутника.

В деяких випадках щільність потоків частинок була досить великою, що призводило до перевищення можливостей апаратури цифрової обробки сигналів приладу СТЕП-Ф [11]. В цих випадках спостерігалися горизонтальні ділянки на кривих реєстрації радіаційних поясів, наприклад, 8 та 11 травня 2009 р. (див. рис. 1).

## 4. Спостереження поясу $S_2/N_2$ за відсутності внутрішнього радіаційного поясу Ван Алена

У зазначений вище часовий період реєстрації даних, в окремі дні травня спостерігались випадки, коли потоки електронів у внутрішньому радіаційному поясі Ван Алена не відрізнялись від фонових потоків, тобто цей пояс не був помітним. Водно-

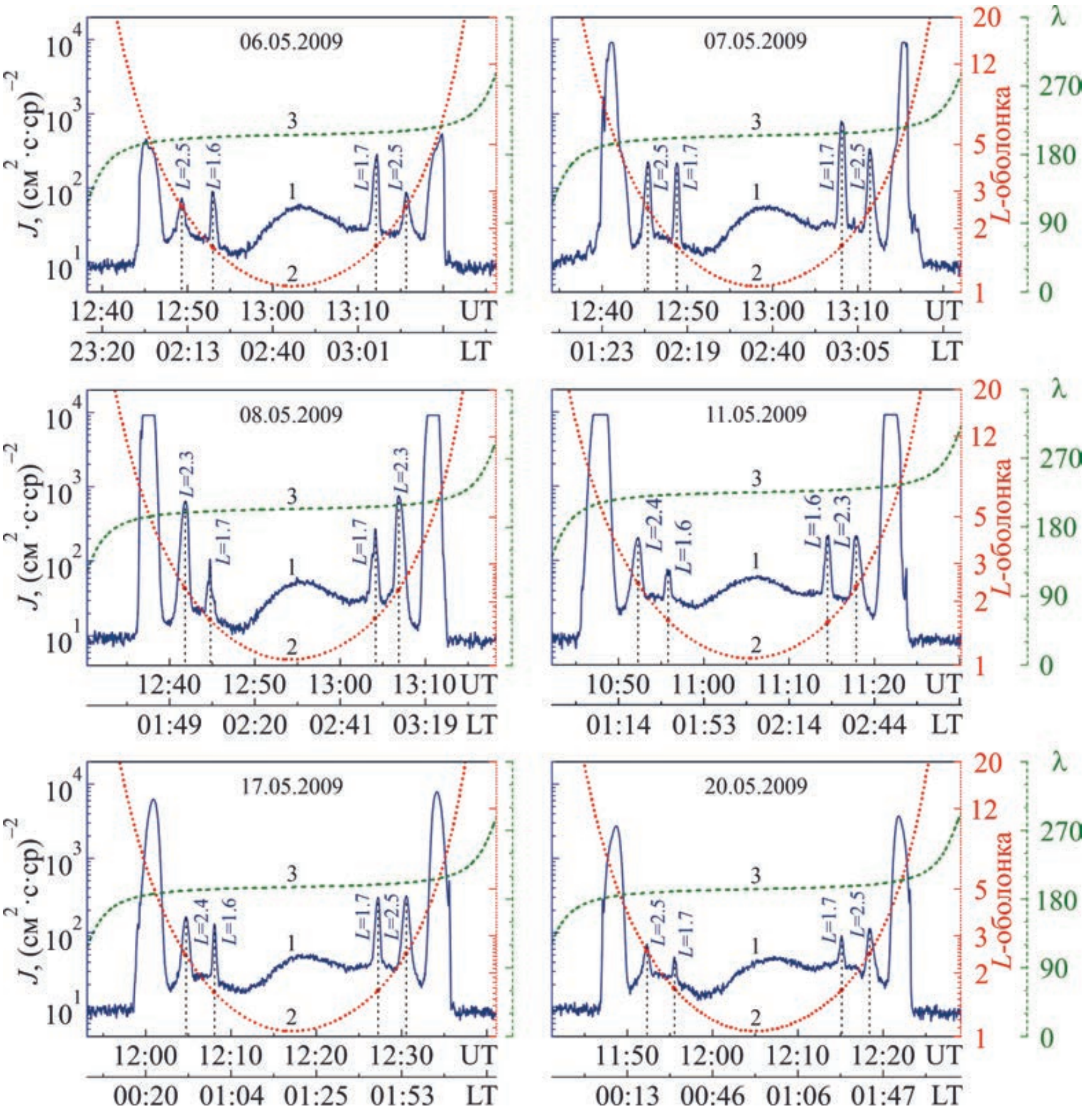

***Рис. 1.*** Типова структура з трьох радіаційних поясів Землі, яка спостерігалась 6, 7, 8, 11, 17, 20 травня 2009 р. в каналі змішаній реєстрації електронів з енергіями $\Delta E_e = 180 \div 510$ кеВ і протонів з енергіями $\Delta E_p = 3.5 \div 3.7$ МеВ вночі за місцевим часом: криві 1 – щільність потоків частинок, пунктирні криві 2 – параметр Мак-Ілвайна $L$, пунктирні криві 3 – географічна довгота $\lambda$ час вузький за $L$-параметром додатковий сталий радіаційний пояс $S_2/N_2$ чітко вирізнявся на тлі фонових потоків частинок, як це можна бачити з рис. 2. Максимальна щільність потоків електронів у додатковому поясі досягала значень потоків у внутрішньому поясі Ван Алена (див. рис. 1).

## 5. Виявлення нестабільного внутрішнього радіаційного поясу $S_{12}/N_{21}$

В окремі дні на обмеженій кількості обертів космічного апарата спостерігалися одночасно два додаткових внутрішніх радіаційних пояси, сталий $S_2/N_2$ і нестабільний, що реєструвався між досліджуваним сталим поясом на $L \approx 1.65$ і внутрішнім поясом Ван Алена на $L \approx 2.52$. Надалі цей радіаційний пояс, що спорадично виникав на певний короткий проміжок часу, позначено як $S_{12}/N_{21}$. В табл. 2 наведено результати спостережень у період з 1 по 31 травня 2009 р. додаткового нестабільного внутрішнього радіаційного поясу Землі $S_{12}/N_{21}$ з зазначенням дати, номера оберту космічного апарата, місцевого і світового часів, $L$-оболонки, на якій реєструвалося максимальне значення потоку електронів, географічних координат і власне значенням максимального потоку. З табл. 2 видно, що усереднені значення $L$-оболонки для поясу $S_{12}/N_{21}$ становлять $L = 2.07 \pm 0.14$ у Південній півкулі і $L = 2.06 \pm 0.14$ у Північній півкулі. Прочерками позначено випадки відсутності підвищених

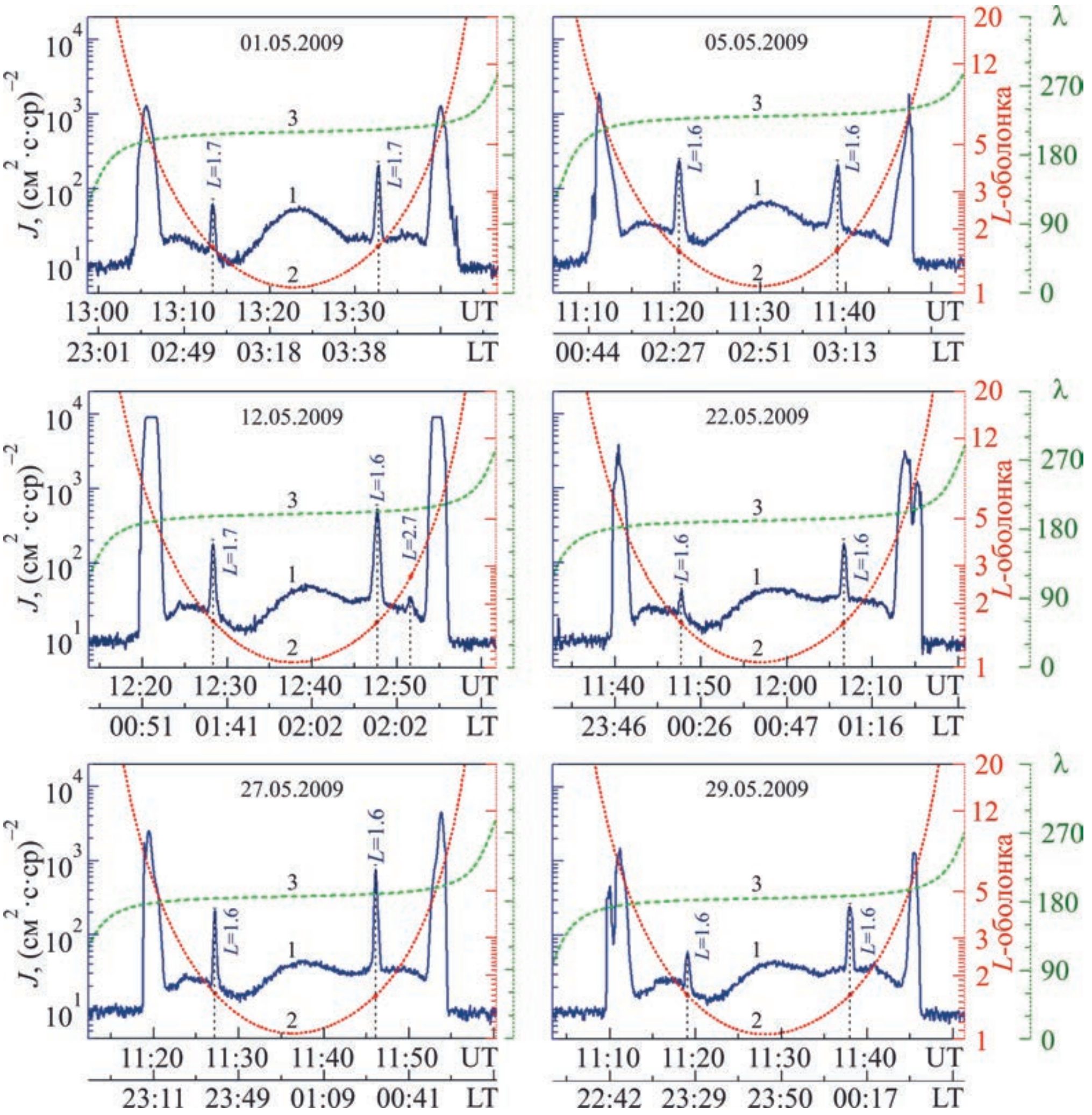

***Рис. 2.*** Типова структура з лише додаткового внутрішнього $S_2/N_2$ та зовнішнього Ван Алена радіаційних поясів Землі, яка спостерігалась 1, 5, 12, 22, 27, 29 травня 2009 р. в каналі змішаної реєстрації електронів з енергіями $\Delta E_e = 180 \div 510$ кеВ і протонів з енергіями $\Delta E_p = 3.5 \div 3.7$ МеВ вночі за місцевим часом: криві 1 – щільність потоків частинок, пунктирні криві 2 – параметр Мак-Ілвайна *L*, пунктирні криві 3 – географічна довгота $\lambda$

(порівняно з оточуючими фоновими) значень потоків у радіаційному поясі $S_{12}/N_{21}$ в одній з півкуль.

На рис. 3 показано приклад реєстрації нестабільного тимчасового радіаційного поясу Землі $S_{12}/N_{21}$ на $L \approx 2.07$ у травні 2009 р., що спостерігався між досліджуваним додатковим внутрішнім радіаційним поясом $S_2/N_2$ на $L \approx 1.65$ і поясом Ван Алена на $L \approx 2.52$.

Рис. 3 демонструє той факт, що у Північній півкулі радіаційні пояси на $L \approx 1.65$, $L \approx 2.07$ і на $L \approx 2.52$ є більш виразними, ніж у Південній, що, з одного боку, може вказувати на зміну кута зору приладу відносно напрямків руху частинок усередині поясів під час перетину космічним апаратом поясів у Південній і Північній півкулях. З іншого боку, відмінність інтенсивностей потоків частинок у різних півкулях може свідчити про анізотропію у напрямках швидкостей електронів усередині всіх поясів, за винятком зовнішнього радіаційного поясу Ван Алена.

## 6. Розподіли інтенсивності електронних потоків за *L*-оболонками

Надання результатів аналізу як залежностей щільності потоків частинок від *L*-оболонки дозволяє зазначити певні нові особливі риси в розподілах субрелятивістських електронів усередині магнітосфери та поза її межами, зважаючи на те, що нахил орбіти космічного апарата ста-

*Таблиця 2.* **Перелік випадків спостереження додаткового нестабільного внутрішнього електронного радіаційного поясу Землі $S_{12}/N_{21}$ у період з 1 по 31 травня 2009 р.**

| Дата | Номер оберту супутника | Час реєстрації поясу $S_{12}$ у Південній півкулі, год:хв:с | | *L*-оболонка | D1e, частинка/(см²·с·ср) | Довгота, ° (0°÷360°) | Широта, ° (–90°÷90°) | Час реєстрації поясу $N_{21}$ у Північній півкулі, год:хв:с | | *L*-оболонка | D1e, частинка/(см²·с·ср) | Довгота, ° (0°÷360°) | Широта, ° (–90°÷90°) |
|---|---|---|---|---|---|---|---|---|---|---|---|---|---|
| | | Місцевий (LT) | Світовий (UT) | | | | | Місцевий (LT) | Світовий (UT) | | | | |
| 03 | 1 | – | – | – | – | – | – | 03:40:28 | 15:03:16 | 2.1 | 28 | 189 | 48 |
| 04 | 2 | 15:43:24 | 22:12:24 | 2.2 | 76 | 263 | –57 | – | – | – | – | – | – |
| 06 | 3 | – | – | – | – | – | – | 03:11:37 | 11:37:46 | 2.2 | 37 | 234 | 42 |
| 07 | 4 | – | – | – | – | – | – | 03:03:47 | 13:09:32 | 1.9 | 43 | 209 | 42 |
| 08 | 5 | 02:04:53 | 14:19:44 | 2.0 | 28 | 176 | –38 | 03:01:16 | 14:42:40 | 2.0 | 82 | 185 | 48 |
| 09 | 6 | 15:02:37 | 21:51:32 | 2.2 | 42 | 258 | –56 | 13:54:41 | 21:26:24 | 2.2 | 63 | 247 | 38 |
| 10 | 7<br>8 | 01:35:58<br>14:46:09 | 09:20:36<br>23:21:40 | 2.1<br>2.1 | 69<br>47 | 244<br>231 | –52<br>–50 | 02:35:14<br>13:41:56 | 09:43:52<br>22:56:44 | 2.0<br>2.1 | 45<br>75 | 253<br>221 | 35<br>43 |
| 11 | 9<br>10 | –<br>14:40:46 | –<br>20:06:24 | –<br>1.8 | –<br>696 | –<br>279 | –<br>–51 | 02:39:34<br>13:46:26 | 14:30:46<br>19:44:54 | 2.1<br>1.9 | 38<br>44 | 182<br>270 | 49<br>29 |
| 14 | 11 | – | – | – | – | – | – | 13:12:43 | 00:18:38 | 2.1 | 51 | 194 | 48 |
| 15 | 12 | – | – | – | – | – | – | 02:01:23 | 12:36:32 | 1.8 | 46 | 201 | 42 |
| 18 | 13 | 13:49:12 | 19:37:50 | 1.9 | 58 | 273 | –54 | – | – | – | – | – | – |
| 23 | 14 | – | – | – | – | – | – | 01:02:17 | 12:04:24 | 2.0 | 42 | 195 | 46 |
| 26 | 15 | 12:50:20 | 19:05:32 | 2.1 | 265 | 266 | –56 | 11:44:37 | 18:41:08 | 2.1 | 47 | 256 | 35 |
| 30 | 16<br>17<br>18 | 23:10:06<br>23:14:07<br>– | 11:12:02<br>12:49:10<br>– | 2.3<br>2.2<br>– | 29<br>61<br>– | 180<br>156<br>– | –42<br>–37<br>– | 00:12:27<br>00:14:14<br>00:15:30 | 11:36:34<br>13:12:46<br>14:48:50 | 2.2<br>2.1<br>2.2 | 63<br>156<br>99 | 189<br>165<br>142 | 50<br>52<br>53 |
| 31 | 19<br>20 | 22:54:50<br>– | 06:18:28<br>– | 1.9<br>– | 97<br>– | 249<br>– | –50<br>– | 23:48:52<br>00:07:17 | 06:40:16<br>13:08:46 | 1.9<br>2.2 | 51<br>45 | 257<br>165 | 31<br>52 |
| | | Усереднене значення *L* 2.07±0.14 | | | | | | Усереднене значення *L* 2.06±0.13 | | | | | |

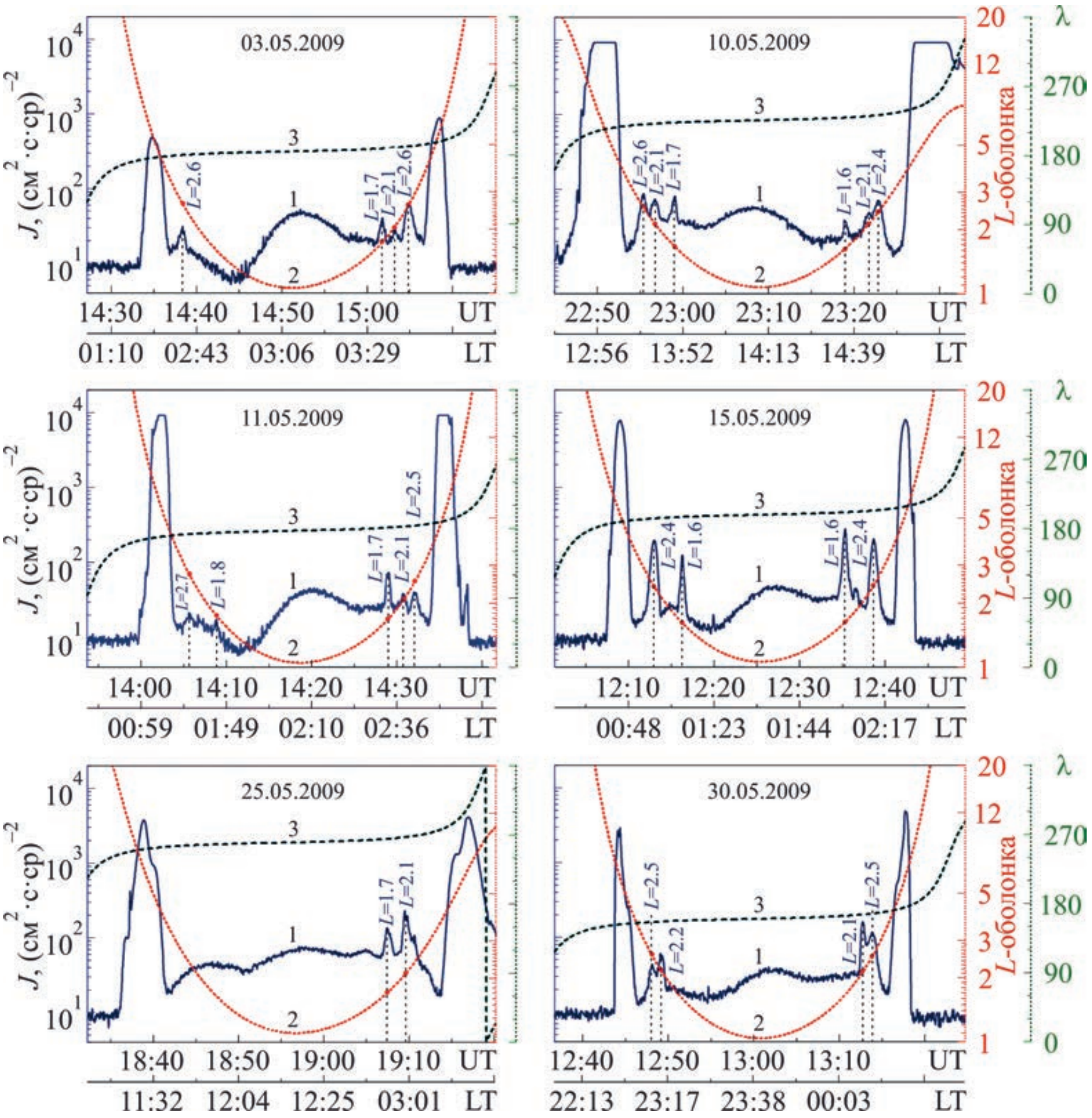

***Рис. 3.*** Приклади спостереження нестабільного радіаційного поясу $S_{12}/N_{21}$ на $L \approx 2.07$ у травні 2009 р.: криві 1 – щільність потоків частинок, пунктирні криві 2 – параметр Мак-Ілвайна *L*, пунктирні криві 3 – географічна довгота $\lambda$ новив близько 84°. На рис. 4 показано розподіли щільності потоків частинок за *L*-оболонками в обох півкулях Землі. Верхні два графіки (записи 7 і 17 травня) демонструють наявність трипоясної структури розподілу частинок усередині магнітосфери. Середні графіки (записи 22 і 29 травня) показують наявність тільки додаткового сталого внутрішнього радіаційного поясу на $L = 1.65$ і зовнішнього радіаційного поясу Ван Алена.

Крім того, на цих двох графіках виявилася ще одна особливість, пов'язана з поведінкою частинок за межами зовнішньої кромки зовнішнього радіаційного поясу Землі. Інколи в Північній або в Південній півкулі поза межами зовнішнього краю зовнішнього радіаційного поясу ($L \geq 7 \div 8$) спостерігається підвищення інтенсивності потоків частинок у широкій смузі *L*-оболонок, що відповідають високоширотній, зокрема авроральній, області квазізахвату високоенергійних заряджених частинок. Підвищені потоки спостерігаються аж до меж головної ударної хвилі земної магнітосфери ($L \approx 10 \div 12$) та за щільністю досягають максимальних значень усередині зовнішнього радіаційного поясу, як це можна бачити з рис. 4. Оскільки в цей час в міжпланетному просторі, на $L \geq 14 \div 20$, не спостерігається підвищення потоків заряджених частинок, тобто відсутні варіюючі потоки сонячних космічних променів або ж посилені потоки частинок, прискорених на міжпланетних ударних хвилях, єдиним джерелом збільшення щільності потоку частинок у перехідному шарі земної магнітосфери можуть бути нестаціонарні

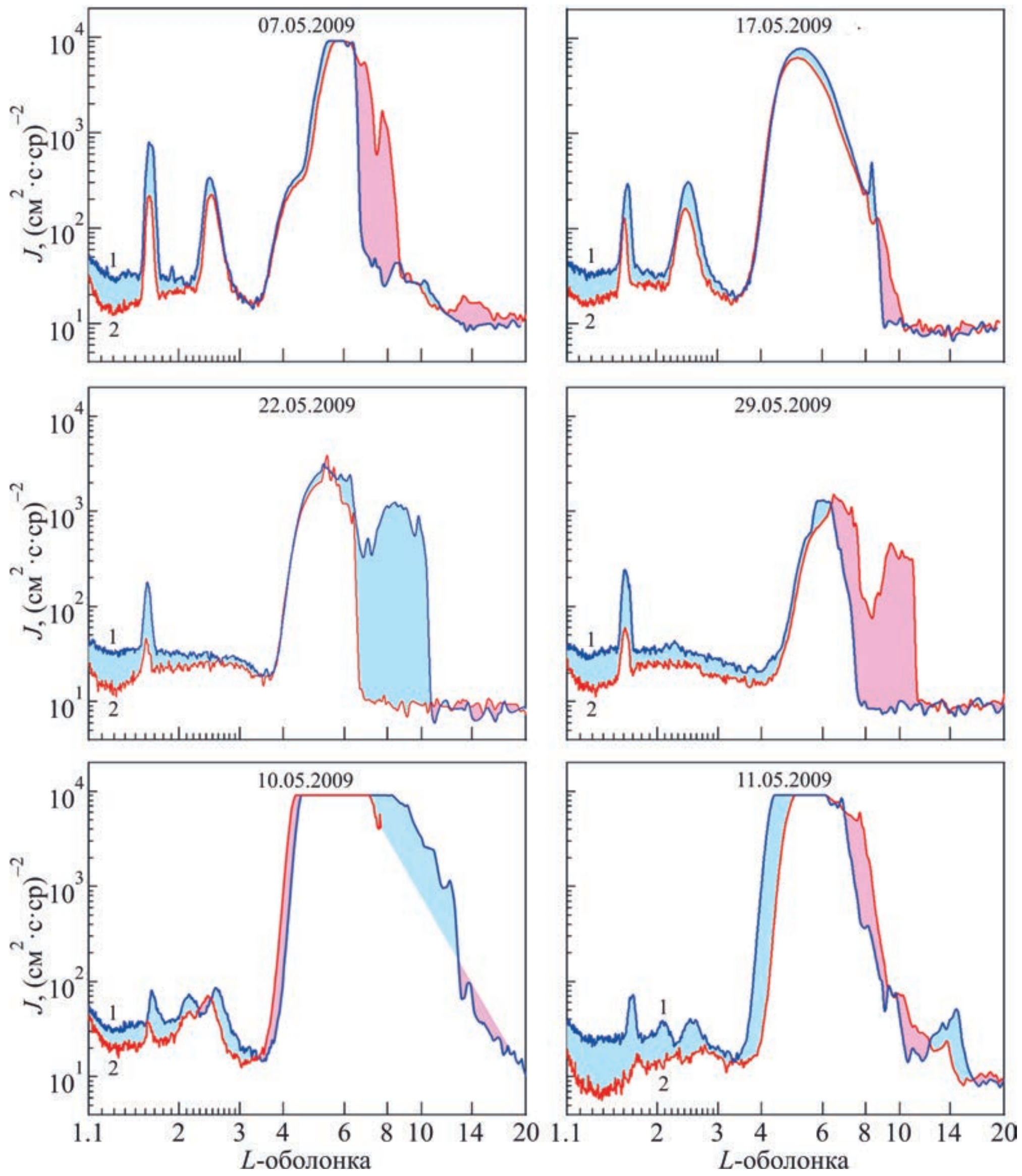

***Рис. 4.*** Розподіли щільності потоків субрелятивістських електронів за *L*-оболонками в Північній (криві 1) і Південній (криві 2) півкулях Землі процеси усередині самої магнітосфери. Такими процесами, зокрема, можуть бути прискорення, пітч-кутова та радіальна дифузії електронів зовнішнього поясу назовні, тобто перехід на вищі дрейфові оболонки геомагнітного поля.

Й нарешті, на нижніх графіках рис. 4 (записи 10 і 11 травня) показано випадки одночасної реєстрації додаткових сталого $S_2/N_2$ і нестабільного $S_{12}/N_{21}$ поясів, внутрішнього і зовнішнього радіаційних поясів Ван Алена. Як можна помітити, ланцюжок внутрішніх поясів виразніше видимий у Північній, аніж у Південній півкулі.

## 7. Проекції пікових значень потоків частинок у внутрішніх радіаційних поясах на географічній мапі Землі

Результати реєстрації сталого додаткового внутрішнього радіаційного поясу $S_2/N_2$, перелічені в табл. 1, та результати реєстрації нестабільного радіаційного поясу $S_{12}/N_{21}$, перелічені в табл. 2, а також розташування внутрішнього радіаційного поясу Ван Алена було нанесено на географічну мапу Землі. На рис. 5 показано проекції позицій реєстрації максимальних значень щільності потоків електронів у внутрішньому радіаційному

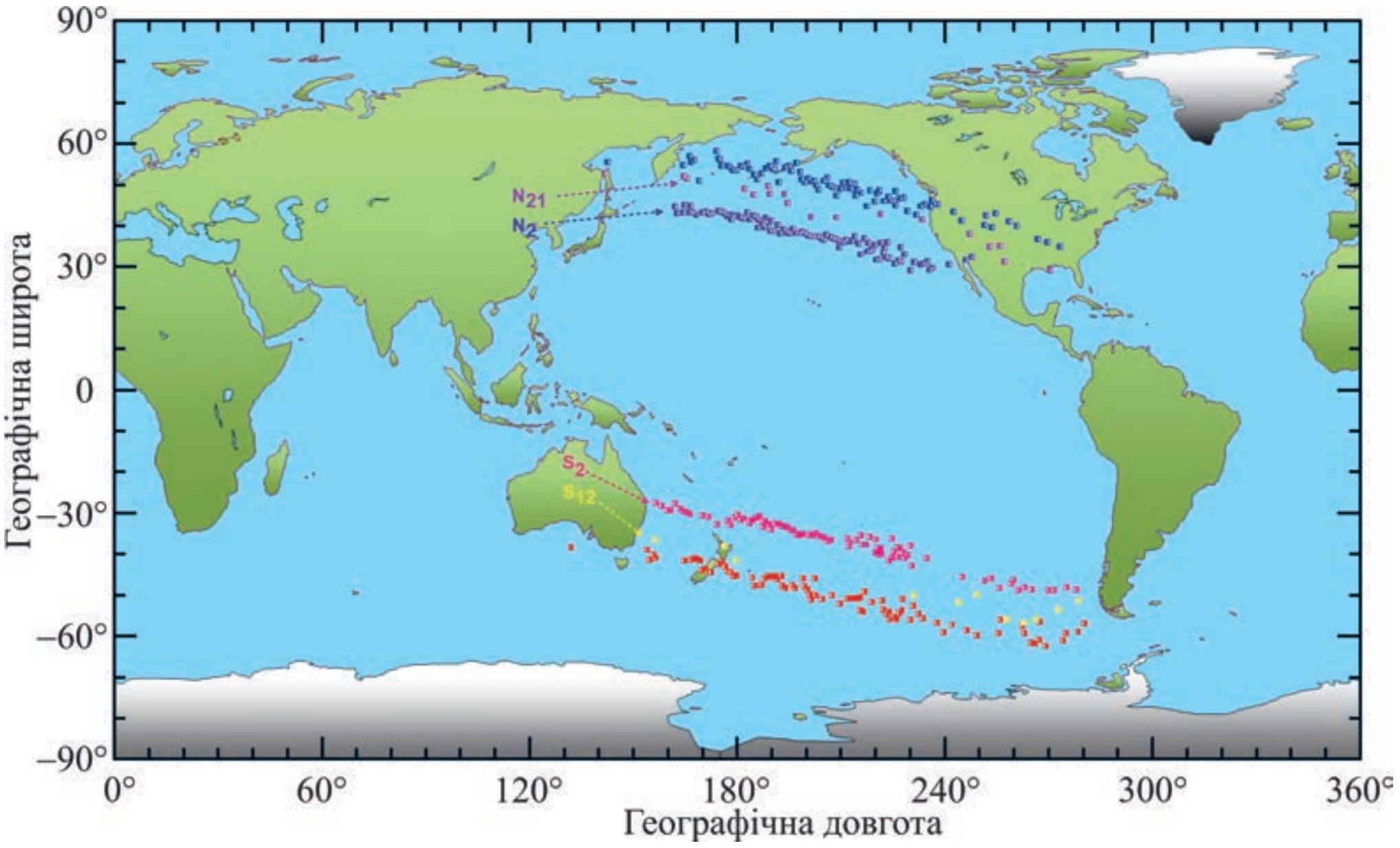

***Рис. 5.*** Проекції позицій реєстрації максимальної щільності потоків частинок у внутрішньому поясі Ван Алена і додаткових внутрішніх радіаційних поясах $S_2/N_2$ і $S_{12}/N_{21}$ на географічній мапі Землі поясі Ван Алена (верхня смуга точок) у Північній півкулі та (нижня смуга точок) у Південній півкулі), у додатковому внутрішньому радіаційному поясі $S_2/N_2$ та у нестабільному додатковому внутрішньому поясі $S_{12}/N_{21}$ у період з 1 по 31 травня 2009 р. З рис. 5 видно, що кількість випадків реєстрації додаткового сталого внутрішнього радіаційного поясу $S_2/N_2$ є не меншою за кількість випадків реєстрації поясу Ван Алена, тоді як темп детектування нестабільного поясу $S_{12}/N_{21}$ є значно меншим. Можна помітити також, що у Північній півкулі кількість випадків реєстрації $N_{21}$ перевищує загальну кількість детектування поясу $S_{12}$ у Південній півкулі.

Додаткові внутрішні радіаційні пояси впевнено реєструються в широкій смузі географічних довгот як на висхідних, так і на низхідних ділянках орбіти супутника, від $\lambda_1 \approx 150°$ до $\lambda_2 \approx 290°$. З урахуванням того, що ПАА спостерігалась приладом СТЕП-Ф у каналі D1e в смузі географічних довгот від $\lambda_{SAA1} \approx 290°$ до $\lambda_{SAA2} \approx 50°$, можна припустити, що виявлені додаткові радіаційні пояси не реєструються на висоті 550 км лише у досить вузькій смузі довгот, що становить $\Delta\lambda \approx \lambda_1 - \lambda_{SAA2} \approx 100°$.

Положення проекцій позицій максимальних значень щільності потоку частинок на географічній мапі Землі, відображене на рис. 5, можна апроксимувати простою лінійною функцією:

$$\varphi = a + b \cdot \lambda,$$

де $\varphi$ – географічна широта. У табл. 3 наведено результати апроксимації лінійною функцією положення максимальних значень щільності потоків електронів внутрішніх радіаційних поясів у Північній і Південній півкулях.

## 8. Висновки

1. Аналіз експериментальних даних з часовою роздільною здатністю $\tau = 2$ с, отриманих зі змішаного каналу реєстрації електронів з енергіями $\Delta E_e = 180 \div 510$ кеВ і протонів з енергіями $\Delta E_p = 3.5 \div 3.7$ МеВ супутникового телескопу електронів і протонів СТЕП-Ф низькоорбітального космічного апарату "КОРОНАС-Фотон", дозволив виявити сталу тришарову структуру електронних радіаційних поясів у земній магнітосфері в період низької сонячної і геомагнітної активності травня 2009 р. Два електронні пояси є відомими з початку космічної ери радіаційними поясами Ван Алена, ще один додатковий сталий шар розташований нижче внутрішнього поясу Ван Алена і формується навколо дрейфової оболонки з параметром Мак-Ілвайна $L = 1.65 \pm 0.05$.

2. Щільність потоків електронів у додатковому сталому внутрішньому радіаційному поясі $S_2/N_2$, який щоденно детектувався на обмеженій

*Таблиця 3.* **Апроксимації положення проекцій позицій максимальних значень густини потоків електронів внутрішніх сталих і нестійких радіаційних поясів на географічній мапі Землі**

| Радіаційний пояс | Лінійна апроксимація |
|---|---|
| Внутрішній Ван Алена (Північна півкуля) | $\varphi = (86.43 \pm 1.40) - (0.18 \pm 0.01) \cdot \lambda$ |
| Додатковий нестійкий $N_{21}$ (Північна півкуля) | $\varphi = (82.60 \pm 2.71) - (0.19 \pm 0.01) \cdot \lambda$ |
| Додатковий сталий $N_2$ (Північна півкуля) | $\varphi = (75.63 \pm 1.12) - (0.19 \pm 0.01) \cdot \lambda$ |
| Додатковий сталий $S_2$ (Південна півкуля) | $\varphi = (1.87 \pm 0.92) - 0.19 \cdot \lambda$ |
| Додатковий нестійкий $S_{12}$ (Південна півкуля) | $\varphi = (-12.32 \pm 4.47) - (0.16 \pm 0.02) \cdot \lambda$ |
| Внутрішній Ван Алена (Південна півкуля) | $\varphi = (-13.39 \pm 1.16) - (0.18 \pm 0.01) \cdot \lambda$ |

кількості обертів, від 1 до 6, космічного апарату, приблизно рівнялася щільності потоку у внутрішньому поясі Ван Алена. Водночас спостерігались неодноразові випадки, коли за відсутністю поясу Ван Алена детектувався лише додатковий пояс $S_2/N_2$.

3. В окремі дні травня 2009 р. на обмеженій кількості обертів космічного апарата спостерігався не один додатковий внутрішній радіаційний пояс $S_2/N_2$, але два одночасно. Один з них реєструвався між досліджуваним сталим поясом на $L \approx 1.65$ і внутрішнім поясом Ван Алена на $L \approx 2.52$. Підвищені потоки частинок у цьому поясі, позначеному як $S_{12}/N_{21}$, гуртуються навколо дрейфової оболонки $L = 2.06 \pm 0.14$. Помічено, що у Північній півкулі кількість випадків реєстрації шару $N_{21}$ більша за кількість детектування поясу $S_{12}$ у Південній півкулі.

4. Додаткові внутрішні радіаційні пояси виразно реєструються в широкій смузі географічних довгот $\lambda$ як на висхідних, так і на низхідних ділянках орбіти супутника, від $\lambda_1 \approx 150°$ до $\lambda_2 \approx 290°$. Враховуючи, що ПАА спостерігалась приладом СТЕП-Ф у каналі D1e в смузі географічних довгот від $\lambda_{SAA1} \approx 290°$ до $\lambda_{SAA2} \approx 50°$, можна припустити, що виявлені додаткові радіаційні пояси не реєструються на висоті 550 км лише у досить вузькій смузі довгот $\Delta\lambda \approx \lambda_1 - \lambda_{SAA2} \approx 100°$.

5. Виявлено, що окремо в Північній або в Південній півкулі, поза межами зовнішнього краю зовнішнього радіаційного поясу, тобто на $L \geq 7 \div 8$, спостерігаються випадки підвищення щільності потоку частинок в широкій смузі *L*-оболонок, що відповідають високоширотній області квазізахвату високоенергійних заряджених частинок. Підвищені значення щільності потоків спостерігаються аж до меж головної ударної хвилі земної магнітосфери $(L \approx 10 \div 12)$ та рівняються максимальним значенням усередині зовнішнього радіаційного поясу. Джерелом суттєвого збільшення щільності потоку частинок у перехідному шарі земної магнітосфери можуть бути нестаціонарні процеси усередині самої магнітосфери, такі як прискорення, пітч-кутова та радіальна дифузії електронів зовнішнього поясу назовні, тобто перехід на вищі дрейфові оболонки геомагнітного поля.

## СПИСОК ЛІТЕРАТУРИ

*O. V. Dudnik and O. V. Yakovlev*

Institute of Radio Astronomy,
National Academy of Sciences of Ukraine,
4, Mystetstv St., Kharkiv, 61002, Ukraine

EVIDENCE OF THE EARTH'S INNER RADIATION BELTS DURING THE LOW SOLAR AND GEOMAGNETIC ACTIVITY OBTAINED WITH THE STEP-F INSTRUMENT

*Purpose:* The subject of research is the spatio-temporal charged particles in the Earth's magnetosphere outside the South Atlantic magnetic Anomaly during the 11-year cycle of solar activity minimum. The work aims at searching for and clarifying the sustained and unstable new spatial zones of enhanced subrelativistic electron fluxes at the altitudes of the low Earth orbit satellites.
*Design/methodology/approach:* Finding and ascertainment of new radiation belts of the Earth were made by using the data analysis from the D1e channel of recording the electrons of energies of $\Delta E_e = 180 - 510$ keV and protons of energies of $\Delta E_p = 3.5 - 3.7$ MeV of the satellite telescope of electrons and protons (STEP-F) aboard the "CORONAS-Photon" Earth low-orbit satellite. For the analysis, the data array with the 2 s time resolution normalized onto the active area of the position-sensitive silicon matrix detector and onto the solid angle of view of the detector head of the instrument was used.
*Findings:* A sustained structure of three electron radiation belts in the Earth's magnetosphere was found at the low solar and geomagnetic activity in May 2009. The two belts are known since the beginning of the space age as the Van Allen radiation belts, another additional permanent layer is formed around the drift shell with the McIlwaine parameter of $L = 1.65 \pm 0.05.$ On some days in May 2009, the new two inner radiation belts were observed simultaneously, one of those latter being recorded between the investigated sustained belt at $L \approx 1.65$ and the Van Allen inner belt at $L \approx 2.52.$ Increased particle fluxes in this unstable belt have been formed with the drift shell $L \approx 2.06 \pm 0.14.$
*Conclusions:* The new found inner radiation belts are recorded in a wide range of geographic longitudes $\lambda,$ both at the ascending and descending nodes of the satellite orbit, from $\lambda_1 \approx 150°$ to $\lambda_2 \approx 290°.$ Separately in the Northern or in the Southern hemispheres, outside the outer edge of the outer radiation belt, at $L \geq 7 - 8,$ there are cases of enhanced particle flux density in wide range of *L*-shells. These shells correspond to the high-latitude region of quasi-trapped energetic charged particles. Increased particle fluxes have been recorded up to the bow shock wave border of the Earth's magnetosphere $(L \approx 10 - 12).$